%Paper: hep-th/9302020
%From: Petr Horava <horava@yukawa.uchicago.edu>
%Date: Fri, 5 Feb 93 16:57:20 CST

\input phyzzx
\overfullrule=0pt
\tolerance=5000
\twelvepoint
\postdisplaypenalty=0
%
%%%%%%%%%%%%%%%%%%%%%%%%%%%%%%%%%%%%%%%%%%%%%%%%%%%%%%%%%%%%%%%%%%%%%%%%%%%
%
\def\binom#1#2{\pmatrix{#1\cr #2\cr}}
\def\ntwo{$N$=$\,2$}
\def\antib#1#2{\{\kern-.3em\{#1,#2\}\kern-.3em\}_{\rm a}}
\def\npb#1#2#3{{\it Nucl.\ Phys.} {\bf B#1} (19#2) #3}
\def\plb#1#2#3{{\it Phys.\ Lett.} {\bf B#1} (19#2) #3}
\def\mpl#1#2#3{{\it Mod.\ Phys.\ Lett.} {\bf A#1} (19#2) #3}
\def\prl#1#2#3{{\it Phys.\ Rev.\ Lett.} {\bf #1} (19#2) #3}
\def\prd#1#2#3{{\it Phys.\ Rev.} {\bf D#1} (19#2) #3}
\def\cmp#1#2#3{{\it Commun.\ Math.\ Phys.} {\bf #1} (19#2) #3}
\def\ijmp#1#2#3{{\it Int.\ J. Mod.\ Phys.} {\bf A#1} (19#2) #3}
\def\der{{\rm d}}     \def\ii{{\rm i}}      \def\e#1{\,{\rm e}^{#1}}
\def\frac#1#2{{#1 \over #2}}

\def\p{\partial}

\def\semi{\subset\kern-1em\times\;}
\def\intz{\int_\Sigma \der^2z\,}
\def\winfty{w_\infty}
\def\winftytop{\winfty^{\rm top}}

\def\CD{{\cal D}}     \def\CL{{\cal L}}     \def\CR{{\cal R}}
\def\CG{{\cal G}}          \def\CO{{\cal O}}
          \def\CQ{{\cal Q}}
\def\CW{{\cal W}}          \def\CK{{\cal K}}
\def\CM{{\cal M}}

      \def\BD{{\bf D}}      \def\BZ{{\bf Z}}
\def\BC{{\bf C}}      
\def\QL{Q_{\rm L}}
\def\im{{\rm Im}\,}   
\def\re{{\rm Re}\,}   

\def\chaptermessage#1{\message{\the\chapternumber . #1}}
\def\endli{\hfill\break}
%
%%%%%%%%%%%%%%%%%%%%%%%%%%%%%%%%%%%%%%%%%%%%%%%%%%%%%%%%%%%%%%%%%%%%%%%%%%%
\pubnum{EFI-92-70}
\date{January 1993}
\titlepage
\vglue-.1in
%%%%%%%%%%%%%%%%%%%%%%%%%%%%%%%%%%%%%%%%%%%%%%%%%%%%%%%%%%%%%%%%%%%%%%%%%%%
\title{Spacetime Diffeomorphisms and Topological $w_\infty$ Symmetry\break
in Two Dimensional Topological String Theory}
%%%%%%%%%%%%%%%%%%%%%%%%%%%%%%%%%%%%%%%%%%%%%%%%%%%%%%%%%%%%%%%%%%%%%%%%%%%
\vglue-.1in
\author{Petr Ho\v rava\foot{e-mail address:
horava@yukawa.uchicago.edu}\foot{Robert R. McCormick Fellow; research also
supported by the NSF under Grant No.\ PHY90-00386; the DOE under Grant
No.\ DEFG02-90ER40560; the Czech Chart 77 Foundation; and the Czech Acad.\
Sci.\ under Grant No.\ 91-11045.}}
% \medskip
\address{\centerline{Enrico Fermi Institute}
\centerline{University of Chicago}
\centerline{5640 South Ellis Avenue}
\centerline{Chicago, IL 60637, USA}}
\bigskip
\abstract
This paper analyzes spacetime symmetries of topological string theory on a
two dimensional torus, and points out that the spacetime geometry of the model
is that of the Batalin-Vilkovisky formalism.  Previously I found an infinite
symmetry algebra in the absolute BRST cohomology of the model.  Here I find
an analog of the BV $\Delta$ operator, and show that it defines a natural
semirelative BRST cohomology.  In the absolute cohomology, the
ghost-number-zero symmetries form the algebra of all infinitesimal spacetime
diffeomorphisms, extended at non-zero ghost numbers to the algebra of all
odd-symplectic diffeomorphisms on a spacetime supermanifold.  In the
semirelative cohomology, the symmetries are reduced to $\winfty$ at ghost
number zero, and to a topologically twisted \ntwo\ $\winfty$ superalgebra
when all ghost numbers are included.  I discuss deformations of the model that
break parts of the spacetime symmetries while preserving the topological BRST
symmetry on the worldsheet.  In the absolute cohomology of the deformed model,
another topological $\winfty$ superalgebra may emerge, while the semirelative
cohomology leads to a bosonic $\winfty$ symmetry.
\endpage
%%%%%%%%%%%%%%%%%%%%%%%%%%%%%%%%%%%%%%%%%%%%%%%%%%%%%%%%%%%%%%%%%%%%%%%%%%%
\REF\oldwitten{E. Witten, \cmp{117}{88}{353}; {\bf 118} (1988) 411}
\REF\wtopgr{E. Witten, \npb{340}{90}{281}}
\REF\verlindes{E. Verlinde and H. Verlinde, \npb{348}{91}{457}}
\REF\wsurp{E. Witten, in: ``Strings '90,'' Proceedings of the Superstring
Workshop at Texas A\&M, March 1990, eds.: R. Arnowitt et al.\ (World
Scientific, Singapore, 1991)}
\REF\kekeli{K. Li, \npb{354}{91}{711, 725}}
\REF\dvv{R. Dijkgraaf, E. Verlinde and H. Verlinde, \npb{348}{91}{435};
{\bf B352} (1991) 59; and in:  ``String Theory and Quantum Gravity''
(World Scientific, Singapore, 1991)}
\REF\wcohom{E. Witten, \ijmp{6}{91}{2775}}
\REF\tgproof{M. Kontsevich, \cmp{147}{92}{1}\endli
E. Witten, {\it Surveys in Diff.\ Geom.} {\bf 1} (1991) 243; ``On the
Kontsevich Model and Other Models of Two Dimensional Quantum Gravity,'' IAS
preprint IASSNS-HEP-91/ (June 1991); ``Algebraic Geometry of Quantum Gravity
in Two Dimensions,'' IAS preprint IASSNS-HEP-91/74 (October 1991)}
\REF\wittenwzw{E. Witten, \npb{371}{92}{191}}
\REF\dijk{R. Dijkgraaf, ``Intersection Theory, Integrable Hierarchies and
Topological Field Theory,'' IAS preprint IASSNS-HEP-91/91 (December 1991)}
\REF\giddings{S.B. Giddings, \plb{268}{91}{17}}
\REF\smolin{L. Smolin, ``Recent Developments in Nonperturbative Quantum
Gravity,'' Syracuse preprint (February 1992)}
\REF\conerev{I.R. Klebanov, in: ``String Theory and Quantum Gravity `91,''
Proceedings of the Trieste Spring School 1991, eds.: J.A. Harvey et al.\
(World Scientific, 1992)\endli
D. Kutasov, {\it ibid.}}
\REF\polyakov{A.M. Polyakov, \mpl{6}{91}{635}; ``Singular States in 2D
Quantum Gravity,'' Princeton preprint PUPT-1289 (September 1991)}
\REF\dasjevicki{S.R. Das and A. Jevicki, \mpl{5}{90}{1639}}
\REF\wblack{E. Witten, \prd{44}{91}{314}}
\REF\wground{E. Witten, \npb{373}{92}{187}}
\REF\klebpol{I.R. Klebanov and A.M. Polyakov, \mpl{6}{91}{3273}}
\REF\wzwie{E. Witten and B. Zwiebach, \npb{377}{92}{55}}
\REF\klebward{I.R. Klebanov, \mpl{7}{92}{723}}
\REF\toptorus{P. Ho\v rava, \npb{386}{92}{383}}
\REF\shapere{A. Giveon and A. Shapere, \npb{386}{92}{43}}
\REF\wtoporb{E. Witten, \prl{61}{88}{670}}
\REF\lgtorus{E. Verlinde and N.P. Warner, \plb{269}{91}{96}}
\REF\distvafa{J. Distler and C. Vafa, \mpl{6}{91}{259}; ``The Penner Model
and D=1 String Theory,'' Princeton preprint PUPT-1212, Carg\`{e}se Workshop
on ``Random Surfaces, Quantum Gravity and Strings,'' May 1990}
\REF\dmp{R. Dijkgraaf, G. Moore and R. Plesser, ``The Partition Function of
2-D String Theory,'' IAS/Yale preprint IASSNS-HEP-92/48=YCTP-P22-92 (August
1992)}
\REF\ajwinfty{J. Avan and A. Jevicki, \plb{266}{91}{35}; {\bf B272} (1991) 17;
\mpl{7}{92}{357}}
\REF\ms{G. Moore and N. Seiberg, \ijmp{7}{91}{2601}}
\REF\danielss{U.H. Danielsson and D.J. Gross, \npb{366}{91}{3}\endli
U.H. Danielsson, \npb{380}{92}{83}}
\REF\polchinski{D. Minic, J. Polchinski and Z. Yang, \npb{369}{92}{324}}
\REF\conewinf{S.R. Das, A. Dhar, G. Mandal and S.R. Wadia, \mpl{7}{92}{71,
937}; \ijmp{7}{92}{5165}\endli
A. Dhar, G. Mandal and S.R. Wadia, ``Classical Fermi Fluid and Geometric
Action for $c=1$,'' IAS/Tata preprint IASSNS-HEP-91/89=TIFR-TH-91/61 (March
1992); ``Non-relativistic Fermions, Coadjoint Orbits of $W_\infty$ and String
Field Theory at $c=1$,'' Tata preprint TIFR-TH-92/40 (June 1992)}
\REF\winftyrev{I. Bars, in: ``Strings '90,''Proceedings of the Superstring
Workshop at Texas A\&M, March 1990, eds.: R. Arnowitt et al.\ (World
Scientific, Singapore, 1991)\endli
C.N. Pope, L.J. Romans and X. Shen, {\it ibid.}}
\REF\winftyrevnew{E. Sezgin, ``Aspects of $W_\infty$ Symmetry,''
Trieste/Texas A\&M preprint IC/91/206=CTP TAMU-9/91 (1991);
``Area-Preserving Diffeomorphisms, $\winfty$ Algebras and $\winfty$ Gravity,''
Texas A\&M preprint CTP-TAMU-13/92 (February 1992)\endli
C.N. Pope, ``Lectures on $W$ Algebras and $W$ Gravity,'' Texas A\&M preprint
CTP TAMU-103/91 (December 1991)\endli
X. Shen, ``$W$-Infinity and String Theory,'' CERN preprint CERN-TH.6404/92
(February 1992)\endli
C.M. Hull, ``Classical and Quantum $W$ Gravity,'' QMC preprint QMW/PH/92/1
(January 1992)}
\REF\highsym{D.J. Gross, \prl{60}{88}{1229}\endli
E. Witten, {\it Phil.\ Trans.\ Roy.\ Soc.} {\bf A329} (1989) 345}
\REF\ellisold{J. Ellis, N.E. Mavromatos and D.V. Nanopoulos,
\plb{267}{91}{465}; {\bf B272} (1991) 261; {\bf B276} (1992) 56;
{\bf B278} (1992) 246; {\bf B284} (1992) 43}
\REF\ellis{J. Ellis, N.E. Mavromatos and D.V. Nanopoulos, \plb{288}{92}{23}}
\REF\winftybh{I. Bakas and E. Kiritsis, \ijmp{7}{92}{339}\endli
F. Yu, \npb{375}{92}{173}\endli
F. Yu and Y.-S. Wu, ``An Infinite Number of Commuting Quantum $\hat W_\infty$
Charges in the $SL(2,R)/U(1)$ Coset Model,'' Utah preprint UU-HEP-92/11 (May
1992)\endli
T. Eguchi, H. Kanno and S.-K. Yang, ``$W_\infty$ Algebra in Two-Dimensional
Black Hole,'' Cambridge preprint NI-92004=DAMTP 92-64 (September 1992)}
\REF\eguchi{T. Eguchi, \mpl{7}{92}{85}}
\REF\aschwarz{A.S. Schwarz, ``Geometry of Batalin-Vilkovisky Quantization,''
``Semiclassical Approximation in Batalin-Vilkovisky Formalism,'' UC Davis
preprints (June and October 1992)}
\REF\wmaster{E. Witten, \mpl{7}{90}{487}}
\REF\henne{M. Henneaux, \npb{ {\rm (}{\it Proc.\ Suppl.}{\rm )}\ 18A}{90}{47};
\plb{282}{92}{372}}
\REF\hull{C.M. Hull, ``${\cal W}$-Geometry,'' QMC preprint QMW-92-6 (November
1992)}
\REF\scatter{D.J. Gross and P. Mende, \plb{197}{87}{129}; \npb{303}{88}{407}}
\REF\vermaster{E. Verlinde, \npb{381}{92}{141}}
\REF\wittenbisft{E. Witten, \prd{46}{92}{5467}; ``Some Computations in
Background Independent Off-Shell String Theory,'' IAS preprint
IASSNS-HEP-92/63 (October 1992)}
\REF\barton{B. Zwiebach, ``Closed String Field Theory: Quantum Action and
the B-V Master Equation,'' IAS preprint IASSNS-HEP-92/41 (June 1992)}
\REF\stasheff{T. Lada and J. Stasheff, ``Introduction to sh-Lie Algebras
for Physicists,'' North Carolina preprint UNC-MATH-92/2 (September 1992)}
\REF\deform{M. Gerstenhaber and S.D. Schack, ``Algebraic Cohomology and
Deformation Theory,'' in: ``Deformation Theory of Algebras and Structures and
Applications,'' eds: M. Hazewinkel and M. Gerstenhaber, NATO ASI Series C247
(Kluwer, 1988)}
\REF\arnold{V.I. Arnol'd, {\it Ann.\ Inst.\ Fourier} {\bf XVI} (1966) 319;
``Mathematical Methods of Classical Mechanics'' (Springer Verlag, 1978)
App.\ 2.K}
\REF\winftytorus{E.G. Floratos and J. Iliopoulos, \plb{201}{88}{237}\endli
I. Antoniadis, P. Ditsas, E. Floratos and J. Iliopoulos, \npb{300}{88}{549}}
\REF\winftop{C.N. Pope, L.J. Romans, E. Sezgin and X. Shen,
\plb{256}{91}{191}}
\REF\nelson{J. Distler and P. Nelson, \prl{66}{91}{1955}; \npb{366}{91}{255};
``Hidden Symmetry in Topological Gravity,'' Penn/Princeton preprint UPR-0463T
(August 1991)\endli
S. Govindarajan, P. Nelson and S.-J. Rey, \npb{365}{91}{633}\endli
S. Govindarajan, P. Nelson and E. Wong, \cmp{147}{92}{253}\endli
E. Wong, ``Recursion Relations in Semirigid Topological Gravity,'' Penn
preprint UPR-0491T (November 1991)}
\REF\elitzur{S. Elitzur, A. Forge and E. Rabinovici, ``On Effective Theories
of Topological Strings,'' CERN/Racah/SISSA preprint CERN-TH.6326
=RI/143/91/11=SISSA/158/91/EP (November 1991)}
\REF\wcswstrings{E. Witten, ``Chern-Simons Gauge Theory as a String Theory,''
IAS preprint IASSNS-HEP-92-45 (July 1992)}
\REF\mukhivafa{S. Mukhi and C. Vafa, ``Two Dimensional Black Hole as a
Topological Coset Model of c=1 String Theory,'' Harvard/Tata preprint
HUTP-93/A002=TIFR/TH/93-01 (January 1993)}
\REF\newtopo{M. Bershadsky, W. Lerche, D. Nemeschansky and N.P. Warner,
``Extended N=2 Superconformal Structure of Gravity and W-Gravity Coupled to
Matter,'' Caltech/CERN/Harvard/USC preprint
CALT-68-1832=CERN-TH.6694/92=HUTP-A061/92=USC-92/021 (October 1992)}
\REF\newbvpapers{B.H. Lian and G.J. Zuckermann, ``New Perspectives on the
BRST-algebraic Structure of String Theory,'' Toronto/Yale preprint (November
1992)\endli
M. Penkava and A.S. Schwarz, ``On Some Algebraic Structures Arising in
String Theory,'' UC Davis preprint (December 1992)\endli
E. Getzler, ``Batalin-Vilkovisky Algebras and Two-Dimensional Topological
Field Theories,'' MIT preprint (December 1992)}

%%%%%%%%%%%%%%%%%%%%%%%%%%%%%%%%%%%%%%%%%%%%%%%%%%%%%%%%%%%%%%%%%%%%%%%%%%%%%%
%
\chapter{Introduction}
\chaptermessage{Introduction}

Diffeomorphism symmetry is one of the most important yet mysterious
symmetries in recent physics.  It underlies classical gravity, and its full
reconciliation with quantum theory still remains a challenging problem.
In this context, considerable interest has been devoted recently to the
conjecturally underlying phase of quantum gravity and string theory
[\oldwitten--\dijk], in which all spacetime diffeomorphisms become manifest
symmetries of the vacuum.  (See also e.g.\ [\giddings,\smolin] for alternative
approaches to the issue.)  One motivation for these studies is the hope that
the topological phase might be more tractable than the usual phase with local
dynamics, which might be eventually obtainable by some symmetry breaking
mechanism.

String theory represents one of the most promising and successful candidates
for establishing links between the topological and dynamical phases of quantum
gravity.  In spacetime dimensions less than two, these two phases of string
theory are actually equivalent [\tgproof,\dijk].  Spacetime dimension two (or
the central charge $c=1$; see e.g.\ [\conerev] for a review) represents the
verge point, at which the topological symmetry (manifest for $c<1$) is no
longer obvious.  The physical spectrum of two dimensional string theory
consists of the massless ``tachyon'' mode representing the center of mass of
the string, and the so-called ``discrete states'' [\polyakov], representing
remnants of its higher excited states.  The tachyon behaves like a
field-theoretical degree of freedom with local dynamics, describable by an
effective spacetime field theory [\dasjevicki].  On the other hand, the
discrete modes have been conjectured to have an underlying topological origin,
possibly in spacetime [\polyakov,\wblack].  They have already proven to play
a crucial role in the model, as they give rise to its ground ring structure
[\wground] and a $\winfty$ symmetry [\wground--\klebward].  Hence, the
dynamical phase of two dimensional string theory represents an interesting
mixture of field-theoretical and non-local degrees of freedom.  This
non-trivial structure raises many questions, such as whether there exists an
underlying phase with just global degrees of freedom, or what happens to the
topological symmetry of $c<1$ string theory as $c\rightarrow 1$ from below.

In this paper I will try to reveal some relevant features of the hypothetical
topological phase of 2-D string theory, by considering a simple model:
topological string theory on a two dimensional toroidal target.  This
program has been initiated in [\toptorus], where I have shown that -- in what
might be called the absolute BRST cohomology of the model -- the non-zero
fundamental group of the target leads to an infinite number of physical
observables, and that these physical observables give rise to an infinite
symmetry algebra.
\foot{Similar results have also been obtained in critical \ntwo\ superstring
theory by Giveon and Shapere [\shapere].  For some earlier work on the
topological torus, see [\wtoporb,\lgtorus]; topological aspects of $c=1$ were
also studied in [\distvafa,\dmp].}
In this paper I study the spacetime structure of the symmetry algebra, and
extend the results of [\toptorus] to a semirelative BRST cohomology.  Among
the main results are the interpretation of the spacetime geometry of the model
in terms of the Batalin-Vilkovisky (BV) geometry, as well as the existence of
$\winfty$ and topologically twisted \ntwo\ $\winfty$ spacetime symmetries
\foot{Here, as in the dynamical phase of 2D string theory [\wground,\wzwie],
the ``spacetime manifold'' of the model is defined as the space parametrized
by generators of the ground ring, and does not a priori coincide with the
target of the sigma model.  See a more detailed discussion of this issue in
\S{3.2}.}
in the semirelative BRST cohomology.

$\winfty$ algebras have played a central role in various approaches to the
dynamical phase of two dimensional string theory [\wground--\klebward,
\ajwinfty--\winftyrevnew].  They behave as unbroken gauge symmetries, and may
represent residua of some underlying stringy gauge symmetry [\highsym].  The
$\winfty$ symmetry may also have other far-reaching physical consequences,
such as quantum coherence restoration in black hole physics via $W$-hair
carried by stringy black holes [\ellisold--\winftybh].  Near the black hole
singularity, where the black hole coset becomes effectively equivalent to a
topological field theory [\eguchi], the $\winfty$ symmetries may get extended
to a topological $\winfty$ [\ellis].  In this paper we will encounter a
spacetime topological $\winfty$ symmetry in a manifestly topological (or, more
exactly, cohomological) two dimensional string theory.

The paper is organized as follows.  In \S{2} I review some basic results of
[\toptorus], such as the structure of observables and the algebra of
symmetries of two dimensional topological string theory, and recall the
structure of its topological ground ring.  In \S{3} I study the spacetime
picture of the model.  We will see that the spacetime manifold $M$ is a two
dimensional torus, dual to the target.  At non-zero ghost numbers, the
spacetime manifold $M$ gets extended to a supermanifold $\CM$ of dimension
$(2|2)$, with a natural odd-symplectic form on it.  In \S{3.1} and \S{3.2} it
is shown that the bosonic part of the symmetry algebra generates all smooth
local diffeomorphisms of the spacetime manifold $M$, while fermionic
generators extend the symmetry algebra to all odd-symplectic diffeomorphisms
on $\CM$.  The odd-symplectic form, generated on spacetime by the symmetry
algebra, induces an anti-bracket structure on the space of physical
observables. This provides first indications that the spacetime geometry of
the model is actually that of the Batalin-Vilkovisky formalism
[\aschwarz--\henne].

The analogy between the spacetime geometry of the model and the BV geometry is
further pursued in \S{4}, where I identify the analog of the $\Delta$
operator, first as a second-order differential operator on spacetime
(\S{4.1}), and then in terms of the worldsheet CFT as a zero mode of the BRST
superpartner of the energy-momentum tensor (\S{4.2}).  This $\Delta$ operator
is shown to define a semirelative BRST cohomology of the model, which is
analyzed from the point of view of spacetime symmetries in \S{4.3}.  In the
semirelative BRST cohomology, the symmetry algrebras of all spacetime
diffeomorphisms and all odd-symplectic diffeomorphisms get reduced to a
$\winfty$ algebra and a topologically twisted \ntwo\ $\winfty$ superalgebra
(denoted by $\winftytop$ henceforth), respectively.  The existence of a
topological $\winfty$ superalgebra in spacetime has interesting consequences,
such as the existence of two BRST charges: One of them is the usual BRST
charge of the worldsheet topological symmetry that defines the physical states
of the model by the standard BRST cohomology condition; the other one emerges
in spacetime, and becomes a part of the topological $\winfty$ superalgebra of
spacetime symmetries.  This interesting similarity between spacetime
and worldsheet symmetries is further extended in \S{4.4}, where it is pointed
out that the model enjoys a topological $\winfty$ algebra not only in
spacetime, but also on the worldsheet.  \S{4.5} briefly points out that the
relationship between the odd-symplectic geometry and the topological $\winfty$
symmetry may shed some light on the issue of $W$ geometry as studied e.g.\ in
[\hull].

In \S{5} I follow a different strategy which also reduces all spacetime
diffeomorphisms to a $\winfty$ algebra.  The basic Lagrangian is deformed by
BRST invariant terms, which breaks a part of the spacetime symmetry algebra as
obtained in the absolute BRST cohomology.  For a particular deformation,
we recover the $\winfty$ symmetry at ghost number zero, while at non-zero
ghost numbers we get a new $\winftytop$ superalgebra.  This new topological
$\winfty$ superalgebra is entangled in an interesting manner with the
$\winftytop$ superalgebra of the semirelative cohomology:  The spacetime
BRST-like charge of the $\winftytop$ superalgebra of the deformed model is
again the Batalin-Vilkovisky $\Delta$ operator that defines in \S{4} the
semirelative BRST cohomology; on the other hand, the conserved charge that
governs the deformation of the model is exactly the BRST-like charge of the
original $\winftytop$ symmetry, as obtained in the semirelative BRST
cohomology.  In \S{5.4} the results of \S{4} and \S{5} are combined together
-- the  BRST invariant deformation of \S{5.1} is shown to break the symmetries
of the {\it semirelative} BRST cohomology from the topological superalgebra
$\winftytop$ to a bosonic $\winfty$.

\S{6} presents some summarizing remarks, as well as possible generalizations
and comparison with other recent results.

%%%%%%%%%%%%%%%%%%%%%%%%%%%%%%%%%%%%%%%%%%%%%%%%%%%%%%%%%%%%%%%%%%%%%%%%%%%%%
%
\chapter{Symmetries in 2-D Topological String Theory}
\chaptermessage{Symmetries in 2-D Topological String Theory}

In this section I review some basic results of [\toptorus], in a form
suitable for the rest of the paper.

Topological strings on a two dimensional toroidal target are described in
conformal coordinates on the worldsheet by the free-field Lagrangian
$$I_0=\frac{1}{\pi}\intz\!\left(\p_z\bar X\p_{\bar z}X-\chi_z\p_{\bar z}\psi -
\bar\chi_{\bar z}\p_z\bar\psi \right) .\eqn\ttlag$$
Here $X,\bar X$ are complex coordinates on the target, $\psi ,\bar\psi$
are the topological ghost fields of spin $(0,0)$, and $\chi\equiv\chi_z\der z$
and $\bar\chi\equiv\bar\chi_{\bar z}\der\bar z$ are the corresponding
anti-ghosts of spins $(1,0)$ and $(0,1)$ respectively.  The theory is
invariant under the BRST symmetry
$$\eqalign{ [Q,X]&=\psi ,\cr \{ Q,\psi\}&=0,\cr \{ Q,\chi_z\}&=\p_z\bar X,\cr}
\qquad
\eqalign{ [Q,\bar X]&=\bar\psi ,\cr \{ Q,\bar\psi\}&=0,\cr
\{ Q,\bar\chi_{\bar z}\}&=\p_{\bar z}X.\cr}\eqn\brstundeformed$$
The classical Lagrangian is also invariant under the $U(1)$ symmetry whose
conserved charge is the ghost number, taking the usual values of plus one on
ghost fields $\psi$ and $\bar\psi$, and minus one on anti-ghosts $\chi$ and
$\bar\chi$.  Here and in what follows, I have integrated out the auxiliary
fields associated with the antighosts, so that the BRST charge is only
nilpotent on shell.

The kinetic term in \ttlag\ contains an imaginary theta term, caused by the
requirement that the Lagrangian be an BRST anticommutator,
$$I_0\equiv \{ Q,\int_\Sigma\Psi \} ,\eqn\brstcommlag$$
for a proper choice of the gauge-fixing fermion $\Psi$.  This exotic form of
the kinetic term in \ttlag\ has several important consequences, such as the
existence of an infinite number of physical states in the model.  This is in
accord with the fact that, if we ``untwist'' the topological sigma model
\ttlag , the resulting \ntwo\ superconformal CFT is not unitary and does not
have to have a finite number of chiral primaries.

%%%%%%%%%%%%%%%%%%%%%%%%%%%%%%%%%%%%%%%%%%%%%%%%%%%%%%%%%%%%%%%%%%%%%%%%%%%%%
\section{Observables and Symmetries}

Already before coupling to topological gravity, the model described by \ttlag\
has a rich structure of BRST-invariant observables, dealt with in detail in
[\toptorus].  Among them, the point-like BRST-invariant observables form a
ring under operator product expansions (OPE); I refer to it as the
``topological ground ring,'' in analogy with the similarly defined ``ground
ring'' of the dynamical phase of 2-D string theory [\wground].
\foot{It is convenient to include into the topological ground ring all
point-like physical observables, not only those of ghost number zero.
In order to be more precise, we would have to distinguish between the
ground ring $\CR$ that consists of only ghost-number-zero observables, and the
extended ground ring $\CR'$, which contains point-like physical observables of
all ghost numbers.  I hope that no confusion will be caused if we call both
$\CR$ and $\CR'$ the ``ground ring.''}

Given a point-like observable $\CO^{(0)}$, one can construct conserved
BRST-invariant charges on the worldsheet via the hierarchy of ``descent
equations'' of the BRST charge:
\smallskip
$$\eqalign{\{ Q,\CO^{(0)}\} &=0,\cr\{ Q,\CO^{(1)}\} &=\der\CO^{(0)},\cr}
\qquad\eqalign{\{ Q,\CO^{(2)}\} &=\der\CO^{(1)},\cr0 &=\der\CO^{(2)}.\cr}
\eqn\descent$$
\smallskip\noindent
The conserved charges are generated by loop integrals of $\CO^{(1)}$,
$$\CQ\equiv \oint_C\CO^{(1)};\eqn\chargesgeneral$$
they are BRST invariant by the descent hierarchy \descent , and act as
symmetries on the topological ground ring.  Worldsheet integrals of
$\CO^{(2)}$,
$$\CW\equiv\int_\Sigma\CO^{(2)},\eqn\twoformsgeneral$$
which are also BRST invariant by \descent\ provided $\Sigma$ is closed,
serve as possible BRST invariant deformations of the basic Lagrangian,
leading to a family of topologically invariant theories.

In our case, the topological ground ring contains two sets of bosonic and
two sets of fermionic observables [\toptorus], each set being parametrized by
winding numbers around the two non-trivial directions on the target.
Following the notation of [\toptorus] I will denote them by
\smallskip
$$\eqalign{O^{(0)}_{m,n}&\equiv\exp\{\ii (mk_a+nk_b)\bar X (z)-\ii (m\bar k_a
+n\bar k_b)X(\bar z)\} ,\cr
P^{(0)}_{m,n}\equiv&\,\psi\, O^{(0)}_{m,n},\qquad
Q^{(0)}_{m,n}\equiv\bar\psi\, O^{(0)}_{m,n},\qquad
R^{(0)}_{m,n}\equiv\psi\bar\psi\, O^{(0)}_{m,n},\cr}
\eqn\pointobs$$
\smallskip\noindent
where $k_a,k_b$ is a (fixed) basis of the lattice that defines the target
torus, and $\bar X(z)$ (resp.\ $\! X(\bar z)$) is the left-moving (resp.$\!$\
right-moving) component of $X$ (resp.\ $\!\bar X$).  Sometimes we will make
use of a specific choice for $k_a ,k_b$; for example, we can take $k_a=R,
\ k_b=R\tau_0$ with $\im\tau_0\neq 0$, so that $R$ measures the overall
scale of the target, while $\tau_0$ is its modulus.

As discussed in [\toptorus] (and in [\wzwie] in the related case of the
dynamical phase of 2-D strings),  BRST descent equations associate a conserved
charge to each point-like observable.  I will denote the charges that
correspond to the observables of \pointobs\ by
$$\eqalign{\CL^a_{m,n}&=\frac{\ii}{2\pi R\,\im\tau_0}
\{\bar\tau_0\oint_CP^{(1)}_{m,n}-\tau_0\oint_CQ^{(1)}_{m,n}\} ,
\cr\CL^b_{m,n}&=\frac{\ii}{2\pi R\,\im\tau_0}\{\oint_CQ^{(1)}_{m,n}-
\oint_CP^{(1)}_{m,n}\} ,\cr
\CQ_{m,n}&=\frac{1}{2\pi}\oint_CO^{(1)}_{m,n},
\qquad\CG_{m,n}=\frac{1}{2\pi}\oint_CR^{(1)}_{m,n}.\cr}
\eqn\chargessymm$$
Explicit worldsheet expressions for the currents entering the right hand side
of \chargessymm\ can be either calculated directly from \descent\ or found in
[\toptorus].  Ghost numbers of $\CL,\CQ$ and $\CG$ are 0, --1 and +1
respectively.  In \chargessymm\ I have chosen specific linear combinations and
normalizations of the charges so that their commutation relations simplify to
the following form,
$$\eqalign{[\CL^a_{m,n},\CL^a_{p,q}]&=(m-p)\CL^a_{m+p,n+q},\cr
[\CL^b_{m,n},\CL^b_{p,q}]&=(n-q)\CL^b_{m+p,n+q},\cr
[\CL^a_{m,n},\CL^b_{p,q}]&=n\CL^a_{m+p,n+q}-p\CL^b_{m+p,n+q},\cr
\{\CQ_{m,n},\CG_{p,q}\}&=-n\CL^a_{m+p,n+q}+m\CL^b_{m+p,n+q},\cr
\{\CQ_{m,n},\CQ_{p,q}\}&=0,\cr}
\ \eqalign{[\CL^a_{m,n},\CG_{p,q}]&=(m-p)\CG_{m+p,n+q},\cr
[\CL^b_{m,n},\CG_{p,q}]&=(n-q)\CG_{m+p,n+q},\cr
[\CL^a_{m,n},\CQ_{p,q}]&=-p\CQ_{m+p,n+q},\cr
[\CL^b_{m,n},\CQ_{p,q}]&=-q\CQ_{m+p,n+q},\cr
\{\CG_{m,n},\CG_{p,q}\}&=0.\cr}
\eqn\communrelunbr$$
This is the infinite-dimensional symmetry algebra in the matter sector of the
topological string theory on a two dimensional torus, as obtained in
[\toptorus].

%%%%%%%%%%%%%%%%%%%%%%%%%%%%%%%%%%%%%%%%%%%%%%%%%%%%%%%%%%%%%%%%%%%%%%%%%%%%%
\section{Coupling to Topological Gravity on the Worldsheet}
The symmetry algebra \communrelunbr\ is generated exclusively by the matter
sector of the model.  In the full-fledged topological string theory, the
matter Lagrangian \ttlag\ has to be coupled to topological gravity on the
worldsheet.  This would produce an infinite tower of gravitational descendants
$\sigma^{(0)}_p\{\CO\} ,\ p=0,1,\ldots$ , built over each observable
$\CO^{(0)}$ of the matter sector.  The gravitational descendants have
essentially the structure of a direct product,
$$\sigma^{(0)}_p\{\CO\}\equiv \phi^p\cdot\CO^{(0)},\eqn\gravdressing$$
where $\phi$ is the bosonic ``ghost for ghost'' [\wtopgr] of the gravitational
sector, carrying ghost number two.  The point-like observables \gravdressing\
enter the descent hierarchy for the total BRST charge of the coupled
gravity-matter system, thus leading to conserved currents
$\sigma^{(1)}_p\{\CO\}$ and two-forms $\sigma^{(2)}_p\{\CO\}$, both BRST
invariant up to exterior derivative terms.  If the orginal sigma-model
observable $\CO^{(0)}$ carries ghost number $q$, the ghost numbers of
$\sigma^{(0)}_p\{\CO\},\ \sigma^{(1)}_p\{\CO\}$ and $\sigma^{(2)}_p\{\CO\}$
are $2p+q,\ 2p+q-1$ and $2p+q-2$ respectively.  Since there are no point-like
BRST-invariant observables of negative ghost numbers in the topological sigma
model, the ghost numbers of the conserved charges
$$\CQ_p\{\CO\} \equiv\oint_C\sigma^{(1)}_p\{\CO\}\eqn\gravcharge$$
with $p>0$ are strictly positive, and the $\CQ_p\{\CO\}$'s do not extend the
ghost-number-zero symmetry algebra.  In this paper we are mainly interested
in symmetries of ghost number zero, to which the gravitational sector does
not contribute; hence, I will continue working on flat worldsheet and
ignoring the topological gravity sector throughout.

Alternatively, we could avoid the discussion of the gravitational descendants
by considering the topological torus as a string theory by itself.  To get the
right ghost number anomaly on the sphere, we can triple the target to a
six-dimensional torus.  This theory would not have to be coupled to
topological gravity on the worldsheet, and may represent a very interesting
example of an exactly solvable string theory.

%%%%%%%%%%%%%%%%%%%%%%%%%%%%%%%%%%%%%%%%%%%%%%%%%%%%%%%%%%%%%%%%%%%%%%%%%%%%%
\section{Canonical Quantization}

The presence of the imaginary theta term in the sigma model Lagrangian \ttlag\
leads to some subtleties in its canonical quantization on worldsheets with
the Minkowski signature.

As noted above, the imaginary theta term makes the theory in the Euclidean
signature non-unitary, allowing for the existence of the infinite number of
physical states.  In the Minkowski worldsheet signature, we will quantize
the model by continuing analytically the original theory both on the
worldsheet and in the spacetime.  As a result of the double analytic
continuation, the theta term in the Lagrangian becomes purely real.  The
theory, however, continues to be non-unitary, as a result of the Minkowski
signature in spacetime.  The reality of the theta term allows us to quantize
directly by standard techniques, while the non-unitarity preserves physicality
of the infinite number of winding modes.

This prescription for handling the imaginary theta term, followed by the
double analytic continuation back to the Euclidean signature on the
worldsheet as well as in spacetime, leads to the following structure of
zero modes,
$$\eqalign{X(z,\bar z)&=x-\ii\left\{\frac{n_1\tau_0+n_2}{2R\,\im\tau_0}+
R(m_1+m_2\tau_0)\right\}\ln z\cr
&\qquad {}-\ii\left\{\frac{n_1\tau_0+n_2}{2R\,\im\tau_0}\right\}\ln\bar z+
{\rm oscillators},\cr}\qquad n_i,m_i\in{\bf Z}.\eqn\oscillquant$$
This result coincides with the expression for the zero modes as mentioned in
[\toptorus], and will be useful later.

The BRST charge $Q$ acts on \oscillquant\ and identifies physical
configurations, as those with $n_1=n_2=0$.  This physicality condition can be
succinctly rewritten as the condition of holomorphicity of \oscillquant ,
$$\bar\p_{\bar z}X(z,\bar z)=0.\eqn\holomorphicbrst$$
This equation, obtained here from the canonical quantization of the model, is
also known from the functional integral formulation: \holomorphicbrst\ defines
instanton configurations of the theory, and the functional integral is by
standard arguments localized on the instanton moduli space.

Note also that the double Minkowski rotation that we have just used in the
canonical quantization of the topological torus is quite reminiscent of the
analytic continuation proposed in the high energy string scattering
[\scatter,\wtoporb,\highsym].  A possible connection between the high energy
behavior of critical strings on one hand and some aspects of string theory on
the topological torus on the other, has been pointed out in [\wtoporb].

%%%%%%%%%%%%%%%%%%%%%%%%%%%%%%%%%%%%%%%%%%%%%%%%%%%%%%%%%%%%%%%%%%%%%%%%%%%%%%
%
\chapter{Spacetime Interpretation of the Symmetry Algebra}
\chaptermessage{Spacetime Interpretation of the Symmetry Algebra}

The commutation relations \communrelunbr\ of the symmetry algebra do not
look, at the first sight, very illuminating.  The main task of this section is
to elucidate the structure of the symmetries in terms of spacetime geometry.

First we discuss what is meant by the spacetime manifold itself.  The
topological ground ring of the model is generated by two bosons of ghost
number zero,
$$a\equiv\exp\{\ii k_a\bar X(z)-\ii\bar k_aX(\bar z)\} ,\qquad
b\equiv\exp\{\ii k_b\bar X(z)-\ii\bar k_bX(\bar z)\} ,\eqn\coordinvtor$$
and two fermions of ghost number one,
$$\Theta_a\equiv \frac{\ii}{2R\,\im\tau_0}(\bar\tau_0\psi -\tau_0\bar\psi  ),
\qquad
\Theta_b\equiv \frac{\ii}{2R\,\im\tau_0}(\bar\psi -\psi ), \eqn\newfermi$$
such that generic elements of the ground ring are of the form
$$a^mb^n\,\Theta_a^p\,\Theta_b^q,\qquad m,n\in \BZ ,\quad p,q\in\{ 0,1\}.
\eqn\genericelement$$
(In \newfermi , analogously as in \chargessymm , I have switched to the
natural basis in the first cohomology group of the target, associated with
the lattice basis chosen, $k_a,k_b$.)

In analogy with the framework of [\wground,\wzwie] the ``spacetime manifold''
of the model is defined as the manifold parametrized by the generators of the
ground ring.  The bosonic part of this manifold, denoted by $M$ from now on,
is spanned by $a$ and $b$; the full supermanifold, paremetrized by $a,b,
\Theta_a$ and $\Theta_b$  will be denoted by $\CM$.  To some extent, this
definition of spacetime may seem rather {\it ad hoc}, but I hope to offer some
arguments below that the definition is in fact quite plausible.  In
particular, the symmetry algebra which has emerged from the BRST descent
equations acquires a natural interpretation as an algebra of spacetime
symmetries, if we use precisely this definition of spacetime.

%%%%%%%%%%%%%%%%%%%%%%%%%%%%%%%%%%%%%%%%%%%%%%%%%%%%%%%%%%%%%%%%%%%%%%%%%%%
\section{Odd-Symplectic Geometry in Spacetime}

By definition, the bosonic generators form a coordinate system on the
spacetime manifold $M$, which is extended by fermionic generators of non-zero
ghost numbers to a supermanifold $\CM$ of dimension $(2|2)$.  This
supermanifold can be endowed with an odd-symplectic form,
$$\omega =\frac{\der a}{a}\wedge\der\Theta_a +\frac{\der b}{b}\wedge\der
\Theta_b .\eqn\oddsymp$$
It has been shown in [\toptorus] that the symmetry algebra \communrelunbr\ of
the topological torus, as obtained from the OPEs of the worldsheet currents,
is equivalent to the algebra of all $\omega$-preserving infinitesimal
diffeomorphisms, and its elements can be represented by vector fields on the
spacetime supermanifold:
\foot{I will not repeat the calculation here, and refer the reader again to
[\toptorus].}
$$\eqalign{\CL^a_{m,n}&=-a^{m+1}b^n\frac{\p}{\p a}+a^mb^n\Theta_a
\left( m\frac{\p}{\p\Theta_a}+n\frac{\p}{\p\Theta_b}\right) ,\cr
\CL^b_{m,n}&=-a^mb^{n+1}\frac{\p}{\p b}+a^mb^n\Theta_b
\left( m\frac{\p}{\p\Theta_a}+n\frac{\p}{\p\Theta_b}\right) ,\cr
\CG_{m,n}&=a^{m+1}b^n\Theta_b\frac{\p}{\p a}-a^mb^{n+1}\Theta_a\frac{\p}{\p b}
+a^mb^n\Theta_a\Theta_b
\left( m\frac{\p}{\p\Theta_a}+n\frac{\p}{\p\Theta_b}\right) ,\cr
\CQ_{m,n}&=a^mb^n
\left( m\frac{\p}{\p\Theta_a}+n\frac{\p}{\p\Theta_b}\right) .\cr}
\eqn\vectrepresent$$
This is one of our central results: The symmetries of the topological string
theory on a torus can be usefully summarized in terms of odd-symplectic
geometry, which is generated naturally in spacetime.

Odd-symplectic geometry has played a central role in the geometric formulation
of the BRST-BV formalism [\aschwarz,\wmaster,\henne].  It has recently become
increasingly important in string theory:  It gathers much of the on-shell
structure of 2-D string theory [\vermaster], is crucial for understanding
prospects for a background independent formulation of open string field theory
[\wittenbisft], and clarifies the structure of general non-polynomial
covariant closed string field theory [\barton].  For a review of some
algebraic structures involved, see [\stasheff].

In our case, the existence of a natural odd-symplectic form on spacetime has
important consequences.  Besides the usual multiplication, defined already by
worldsheet OPEs, the space of physical observables carries an anti-bracket
multiplication defined by the odd-symplectic form $\omega$.  To avoid possible
confusion and distinguish the anti-bracket from the anti-commutator, I will
denote the anti-bracket by $\antib\ \ $.  The explicit definition of
$\antib\ \ $ on two elements $f,g$ of the ground ring is
$$\antib fg=\omega^{-1}(\der f,\der g).\eqn\antibrackongr$$
Here d is the exterior derivative on $\CM$ defined by
$$\der f=\frac{\p f}{\p a}\der a+(-1)^{|f|}\frac{\p f}{\p\Theta_a}\der\Theta_a
+\frac{\p f}{\p b}\der b+(-1)^{|f|}\frac{\p f}{\p\Theta_b}\der\Theta_b,
\eqn\extsuperder$$
$\omega^{-1}$ is the inverted odd-symplectic form,
$$\omega^{-1}\equiv a\frac{\p}{\p a}\wedge\frac{\p}{\p\Theta_a}+
b\frac{\p}{\p b}\wedge\frac{\p}{\p\Theta_b},\eqn\invertedomega$$
and $|f|$ denotes the ghost number of $f$.  In the natural coordinate system
given by $a,b,\Theta_a,\Theta_b$, the anti-bracket acquires the following
form,
$$\antib fg=(-1)^{|f|}\left(a\frac{\p f}{\p a}\frac{\p g}{\p\Theta_a}+
b\frac{\p f}{\p b}\frac{\p g}{\p\Theta_b}\right) +(-1)^{(|f|+1)\cdot |g|}
\left( a\frac{\p g}{\p a}\frac{\p f}{\p\Theta_a}+b\frac{\p g}{\p b}
\frac{\p f}{\p\Theta_b}\right) .\eqn\antibrackcoord$$
The anti-bracket \antibrackongr\ defines on the space of physical observables
the algebraic structure of what is called by mathematicians a G-algebra (or a
Gerstenhaber algebra; cf.\ [\deform] and references therein).

Having understood the structure of the full symmetry algebra in terms of
the odd-symplectic geometry in spacetime, we can also clarify the structure of
the bosonic, ghost-number-zero symmetries.  Indeed, the bosonic part of the
symmetry algebra \vectrepresent , which can be obtained simply by ignoring all
$\Theta_a,\Theta_b$-dependent terms in \vectrepresent , consists of all formal
diffeomorphisms of the spacetime manifold $M$:
$$\left.\CL^a_{m,n}\frac{}{}\right|_{\Theta=0}=-a^{m+1}b^n\frac{\p}{\p a},
\qquad\left.\CL^b_{m,n}\frac{}{}\right|_{\Theta=0}=-a^mb^{n+1}\frac{\p}{\p b}.
\eqn\vectdiffonm$$
The term ``formal'' refers here to the fact that the coefficients of the
vector fields \vectdiffonm\ are formal polynomials in the bosonic ground-ring
generators, so that we are still in the realm of algebraic geometry.  When we
switch from algebraic geometry to differential geometry in the following
subsection, we will find out why the anti-bracket is naturally defined on the
space of physical observables, or in other words, why $\omega$ exists on
$\CM$.  We will also see that the Gerstenhaber algebra of physical observables
is a very special one, known to mathematicians since the sixties.

%%%%%%%%%%%%%%%%%%%%%%%%%%%%%%%%%%%%%%%%%%%%%%%%%%%%%%%%%%%%%%%%%%%%%%%%%%%%%
\section{Spacetime versus Target}

In the dynamical phase of two dimensional string theory, the relation between
the target of the Liouville approach and the spacetime of the matrix model is
highly nontrivial.  Let us recall that a natural set of coordinates on the
spacetime of the usual phase, defined again as the manifold spanned by the
generators of the ground ring [\wground,\wzwie], is given by a time
coordinate, the matrix-model eigenvalue $\lambda$, and its conjugate
momentum.  As has been argued e.g.\ in [\ms], the eigenvalue $\lambda$ and the
target dimension may be related by a complicated integral transform.

On the other hand, in the topological theory that we are studying here, the
relationship between the target (parametrized by $X$ and $\bar X$ of \ttlag )
and the spacetime (parametrized by the bosonic generators of the ground ring,
$a$ and $b$) simplifies considerably.  First of all, we can see e.g.\ from
the form of the preserved symplectic form $\omega$ that a convenient set of
spacetime coordinates is given by the logarithms of $a$ and $b$,
$$A\equiv -\ii\ln a,\qquad B\equiv -\ii\ln b,\eqn\logcoords$$
rather than by $a$ and $b$ themselves.  The new coordiates can also be
represented in terms of the field content of the underlying CFT on the
worldsheet, as specific linear combinations of $X(\bar z)$ and $\bar X(z)$.
Making use of the results of \S{2.3} on the hamiltonian formulation, we obtain
the following mode expansion for the spacetime coordinates $A$ and $B$:
$$\eqalign{A&=u+\frac{n_1\,\re\tau_0+n_2}{\im\tau_0}\tau +n_1\sigma +
{\rm oscillators,}\cr
B&=v+\frac{n_1\,|\tau_0|^2+n_2\,\re\tau_0}{\im\tau_0}\tau -n_2\sigma
+{\rm oscillators,}\cr}
\quad n_1,n_2\in\BZ .\eqn\logcoordsmodes$$
As we can see from this expansion, the new set of spacetime coordinates
parametrize a torus, which is -- as far as its metric properties are concerned
-- dual to the target.  In the topological string theory on a torus, the
spacetime and the target are thus dual to each other.

This fact has one immediate consequence:  We can interpret the ground ring
\genericelement\ as the ring of all smooth functions, and the symmetry algebra
\vectrepresent\ as the algebra generating all local, smooth diffeomorphisms on
the spacetime manifold -- the algebraic geometry of the ground ring turned
into the differential geometry of the spacetime torus.  The natural emergence
of all smooth spacetime functions in our topological theory can be contrasted
with the dynamical phase of two dimensional string theory, where the natural
functions on the spacetime quadric [\wground] are given by polynomial
functions on one hand, and by origin-supported distributions on the other.  In
the topological phase, spacetime geometry has been ``smoothed out'' when
compared to the dynamical phase.

With this interpretation of the spacetime manifold, we can easily understand
the structure of its fermionic extension $\CM$.  The fermionic generators of
the ground ring, $\Theta_a$ and $\Theta_b$, are linear combinations of
worldsheet ghosts $\psi ,\bar\psi$ of the topological BRST symmetry, and are
thus one-forms on the target.  In view of the duality between the target and
the spacetime manifold $M$, $\Theta_a$ and $\Theta_b$ become effectively
vector fields on $M$; this fact can also be confirmed directly, by computing
the transformation relations of $\Theta_a,\Theta_b$ under the generators of
spacetime diffeomorphisms, $\CL^a_{m,n}, \CL^b_{m,n}$:
$$\eqalign{\CL^a_{m,n}\cdot\Theta_a&=ma^mb^n\Theta_a,\cr
\CL^b_{m,n}\cdot\Theta_a&=ma^mb^n\Theta_b,\cr}
\qquad\eqalign{\CL^a_{m,n}\cdot\Theta_b&=na^mb^n\Theta_a,\cr
\CL^b_{m,n}\cdot\Theta_b&=na^mb^n\Theta_b.\cr}\eqn\actonfermi$$
The supersymmetric extension $\CM$ of the spacetime manifold $M$, parametrized
by $A,B,\Theta_a$ and $\Theta_b$, can thus be identified with the tangent
bundle to $M$, with fibers treated as odd dimensions:
$$\CM\equiv \Pi (TM).\eqn\oddbundle$$
The ground ring can be identified with the algebra of all smooth multi-vector
fields on $M$.  The Lie bracket that always exists on vector fields induces an
odd-symplectic form (and consequently an anti-bracket) on $\CM$.  This
explains the existence of the anti-bracket $\antib\ \ $.  Together with
$\antib\ \ $, the ground ring becomes the Gerstenhaber algebra of multi-vector
fields on a manifold, one of the first Gerstenhaber algebras ever studied.
The full symmetry algebra of the model consists of all smooth diffeomorphisms
that preserve the anti-bracket.

%%%%%%%%%%%%%%%%%%%%%%%%%%%%%%%%%%%%%%%%%%%%%%%%%%%%%%%%%%%%%%%%%%%%%%%%%%%%%%
\section{Chiral BRST Cohomology and Its Spacetime Intepretation}

Before pursuing further the analogy between the geometry of two dimensional
topological string theory and the geometry of the BV formalism, I will briefly
discuss a simple refinement of the previous results, coming from the
possiblity to split the BRST cohomology into chiral sectors on the
worldsheet.  This splitting makes a significant use of the conformal structure
on the worldsheet, and most of its consequences can be lost when the
Lagrangian is deformed by BRST invariant terms that do not preserve worldsheet
conformal invariance.

The chiral splitting on the worldsheet causes a chiral splitting of the
spacetime symmetry structure.  To see this, we will consider the left-moving
sector of the theory, with the chiral BRST charge given by
$$\eqalign{[\QL ,X(z)]&=\psi (z),\cr\{\QL,\psi (z)\}&=0,\cr}
\qquad\eqalign{[\QL ,\bar X(z)]&=0,\cr\{\QL,\chi_z (z)\}&=\p_z\bar X(z).\cr}
\eqn\chiralbrst$$
The space of point-like physical observables of the chiral BRST charge $\QL$
contains one bosonic series of observables with ghost number zero, and one
series of fermionic observables with ghost number one, each of them
parametrized by two winding numbers $m,n$.  We will denote these observables
by
$$O^{(0)}_{{\rm L};m,n}\equiv\e{\ii k_{m,n}\bar X(z)},\qquad
P^{(0)}_{{\rm L};m,n}\equiv\psi\e{\ii k_{m,n}\bar X(z)}.\eqn\chiralobs$$
Here $k_{m,n}\equiv mk_a+nk_b$ is again expressed with the use of a
specific basis $k_a,k_b$ of the target torus.

The observables listed in \chiralobs\ form the chiral ground ring of the
theory.  The chiral ground ring is generated by two bosonic observables
$a_{\rm L}\equiv\e{k_a\bar X(z)},b_{\rm L}\equiv\e{k_b\bar X(z)}$, and one
complex fermionic observable, $\theta\equiv\psi$.  BRST descent hierarchy
of the chiral BRST charge $\QL$ associates with each element of the chiral
ground ring its conserved charge,
$$\CL_{\rm L;m,n}\equiv\oint_C O^{(1)}_{{\rm L};m,n},\qquad
\CQ_{{\rm L};m,n}\equiv\oint_C P^{(1)}_{{\rm L};m,n}.\eqn\chiralch$$
These charges act as vector fields on the space parametrized by the generators
of the ground ring.  We can see that the chiral symmetry algebra contains
just two series of conserved charges, one bosonic of ghost number one and
one fermionic of ghost number minus one.

In the full theory, the left-movers are coupled to right-movers by matching
left-moving and right-moving winding numbers, which leads to identification
$$a\equiv a_{\rm L}\cdot a_{\rm R},\qquad
b\equiv b_{\rm L}\cdot b_{\rm R},\eqn\chiraldecomp$$
and recovers the non-chiral structure of the spacetime manifold as studied
in previous subsections.
\endpage
%%%%%%%%%%%%%%%%%%%%%%%%%%%%%%%%%%%%%%%%%%%%%%%%%%%%%%%%%%%%%%%%%%%%%%%%%%%%%%
%
\chapter{Spacetime Topological Symmetry and $\winfty$}
\chaptermessage{Spacetime Topological Symmetry and W-Infinity}

In the usual phase of two dimensional string theory, the BRST descent
equations lead -- when applied to the discrete states -- to symmetry algebras
of a $\winfty$ type.  Thus far, in the topological theory we have studied, the
ghost-number-zero symmetry algebra that naturally emerged was the algebra of
all infinitesimal spacetime diffeomorphisms, ${\rm Diff}_0(T^2)$.  In this
section (and in \S{5}) we find a mechanism that reduces the algebra of all
spacetime diffeomorphisms to a $\winfty$ algebra, thus establishing some
contact between the topological and physical theories.

%%%%%%%%%%%%%%%%%%%%%%%%%%%%%%%%%%%%%%%%%%%%%%%%%%%%%%%%%%%%%%%%%%%%%%%%%%%%%%
\section{More of the BV Geometry}

The full symmetry algebra \communrelunbr\ of the model consists of all
inifinitesimal diffeomorphisms of the spacetime supermanifold that preserve
the odd-symplectic form $\omega$.  As we have seen, this odd-symplectic form
defines an anti-bracket structure on the space of all smooth spacetime
functions (i.e.\ on the space of physical states), by
$$\antib fg=(-1)^{|f|}\left(a\frac{\p f}{\p a}\frac{\p g}{\p\Theta_a}+
b\frac{\p f}{\p b}\frac{\p g}{\p\Theta_b}\right) +(-1)^{(|f|+1)\cdot |g|}
\left( a\frac{\p g}{\p a}\frac{\p f}{\p\Theta_a}+b\frac{\p g}{\p b}
\frac{\p f}{\p\Theta_b}\right) .\eqn\antibagain$$
It is natural to wonder whether this structure can be completed to the full
geometry of the BV formalism, which contains, besides the anti-bracket defined
by the odd-symplectic form, a nilpotent second-order differential operator
$\Delta$.

A natural candidate for the $\Delta$ operator is
$$\Delta =a\frac{\p^2}{\p a\,\p\Theta_a}+b\frac{\p^2}{\p b\,\p\Theta_b},
\eqn\deltaop$$
which is clearly nilpotent, and generates the anti-bracket \antibagain\
by
$$\antib fg=\Delta (fg) -\Delta (f)\cdot g-(-1)^{|f|}f\cdot\Delta (g).
\eqn\antidelta$$
Recalling that $\Theta_a,\Theta_b$ are essentially the (odd) coordinates along
the fibres of $TM$, the $\Delta$ operator of \deltaop\ is nothing but the
exterior derivative $\der$ on the spacetime manifold, rewritten in the dual
form with the use of a volume element on $M$.  This is in accord with the
cohomology of the nilpotent operator $\Delta$ on the extended ground ring.  It
turns out that the cohomology is equal to a free module over $\BC$, generated
by the two fermionic generators of ghost number one:
$$H(\Delta )\equiv {\rm Ker}\, \Delta /{\rm Image}\, \Delta =
\BC [\Theta_a,\Theta_b].
\eqn\cohostbrst$$
With the interpretation of $\Theta_a,\Theta_b$ as one-forms on the target,
this result is exactly what one would expect from the exterior derivative on
spacetime.

%%%%%%%%%%%%%%%%%%%%%%%%%%%%%%%%%%%%%%%%%%%%%%%%%%%%%%%%%%%%%%%%%%%%%%%%%%%%%
\section{Semirelative BRST Cohomology}

By construction, the model we have studied so far is topological on the
worldsheet, i.e.\ its worldsheet Virasoro symmetries become a part of the
topologically twisted \ntwo\ Virasoro superalgebra.  This topological
algebra, whose commutation relations are
$$\eqalign{\eqalign{[L_m,L_n]&=(m-n)L_{m+n},\cr[L_m,G_n]&=(m-n)G_{m+n},\cr
[L_m,Q_n]&=-nQ_{m+n},\cr}
\qquad\eqalign{[J_m,J_n]&=m\delta_{m+n,0},\cr [J_m,G_n]&=-G_{m+n},\cr
[J_m,Q_n]&=Q_{m+n},\cr}\cr
\qquad\eqalign{\frac{}{}
\{ G_m,Q_n\}&=L_{m+n}+nJ_{m+n}-\half m(m+1)\delta_{m+n,0},\cr
[L_m,J_n]&=-nJ_{m+n}-\half m(m+1)\delta_{m+n},\cr}\cr}
\eqn\wstopovirasoro$$
is realized by the following currents,
$$\eqalign{T_{zz}(z)&\equiv \sum_m\frac{L_m}{z^{m+2}}=-\p_zX\p_z\bar X+
\chi_z\p_z\psi,\cr G_{zz}(z)&\equiv\sum_m\frac{G_m}{z^{m+2}}=-\chi_z\p_zX,\cr}
\qquad\eqalign{Q_z(z)&\equiv\sum_m\frac{Q_m}{z^{m+1}}=-\psi\p_z\bar X,\cr
J_z(z)&\equiv\sum_m\frac{J_m}{z^{m+1}}=\psi\chi_z,\cr}
\eqn\wsviracurrents$$
and analogously for the right-movers $\bar L_m,\bar G_m,\bar Q_m,\bar J_m$.
The total BRST charge $Q$ of the model is given by the sum of the zero modes
of the ghost-number-one scalar current, $Q\equiv Q_0+\bar Q_0$.  Physical
states of the topological sigma model have been defined by the (absolute) BRST
cohomology of $Q$, i.e.\ $Q|{\rm phys}\rangle =0$ and $|{\rm phys}\rangle
\sim |{\rm phys}\rangle +|\Lambda\rangle$, with $\Lambda$ arbitrary.

Now imagine that we are interested in a direct spacetime description of the
model, in terms of a string field theory.
\foot{Strictly speaking, we should treat the model as a part of a full string
theory, either by coupling it to topological gravity or by introducing two
other complex dimensions.  In what follows, I implicitly assume that the model
is a part of such a complete theory.}
The quadratic part of the string field action is supposed to reproduce the
BRST cohomology condition,
$$Q|\Psi\rangle =0,\eqn\eomsft$$
as the equation of motion for the string field $\Psi$, which is an element of
the Hilbert space of the model.  By simple ghost-number counting, we can see
that the naive kinetic term $\langle\Psi |Q|\Psi\rangle$ of the string-field
action requires an additional insertion of a ghost-number-one fermionic
operator $C$.  The kinetic term of the string-field action becomes
$$S_2=\langle\Psi |CQ|\Psi\rangle .\eqn\sftaction$$

In critical string theory, the role of $C$ is played by the zero mode $c^-_0$
of the diffeomorphism ghosts [\barton].  By standard arguments, reviewed e.g.\
in [\barton,\wcswstrings], it is crucial for the gauge invariance of the
string field action \sftaction\ that the physical states belong to an
appropriate semirelative BRST cohomology, defined as the cohomology of $Q$
equivariant with respect to the charge $B$ conjugated to $C$.  Notably, the
same charge $B$ that enters the semirelative cohomology condition, enjoys
another important role in the model -- it makes a zero mode of the worldsheet
energy-momentum tensor an exact BRST commutator:
$$\{ Q,B\}=L_0-\bar L_0.\eqn\exactenmom$$
Whereas in topological string theory it may be intricate to identify the
proper candidate for $C$ (see [\wcswstrings]), we have an excellent candidate
for $B$:  The zero mode $G_0^-\equiv G_0-\bar G_0$ of the fermionic
$Q$-superpartner of $T(z)-\bar T(\bar z)$.

Anticipating this structure even before completing the string theory, we can
define the semirelative BRST cohomology of the topological sigma model as the
cohomology of $Q$, equivariant with respect to $G^-_0$:
$$\eqalign{Q|{\rm phys}\rangle&=0,\cr G^-_0|{\rm phys}\rangle&=0,\cr}\qquad
\eqalign{|{\rm phys}\rangle &\sim |{\rm phys}\rangle +Q|\Lambda\rangle ,\cr
G^-_0|\Lambda\rangle&=0.\cr}\eqn\brstequiv$$
It is straightforward to demonstrate with the use of the explicit expressions
for the worldsheet currents \wsviracurrents\ that the action of $G^-_0$ on the
worldsheet fields coincides with that of the searched-for $\Delta$ operator of
\deltaop , so we can identify
$$G^-_0=\Delta .\eqn\geeeqdelta$$

In \S{4.1} we have constructed the fermionic nilpotent operator $\Delta$ as
the object that completes the odd-symplectic geometry in spacetime to the
geometry of the BV formalism.   Now we have found the worldsheet
representation of $\Delta$ and have shown that it defines a natural
semirelative BRST cohomology condition \brstequiv .

%%%%%%%%%%%%%%%%%%%%%%%%%%%%%%%%%%%%%%%%%%%%%%%%%%%%%%%%%%%%%%%%%%%%%%%%%%%%
\section{Spacetime Symmetries in the Semirelative BRST Cohomology}

To identify the structure of symmetries in the semirelative BRST cohomology,
consider first the subspace $\CK$ of the ground ring that consists of all
elements annihilated by $\Delta$, i.e.\ $\CK\equiv{\rm Ker}\,\Delta$.
Explicitly, the action of $\Delta$ on the basis of the ground ring is given by
$$\eqalign{\Delta (a^mb^n)&=0,\cr
\Delta (a^mb^n\Theta_a)&=ma^mb^n,\cr
\Delta (a^mb^n\Theta_b)&=na^mb^n,\cr
\Delta (a^mb^n\Theta_a\Theta_b)&= a^mb^n(m\Theta_b-n\Theta_a).\cr}
\eqn\semirdelatact$$
Consequently, the space $\CK$ of modes annihilated by $\Delta$ is spanned by
$$1,\quad a^mb^n, \quad a^mb^n(n\Theta_a+m\Theta_b),\quad\Theta_a,\quad
\Theta_b,\quad\Theta_a\Theta_b.\eqn\genkerdelta$$
These point-like observables produce, via the BRST descent equations, a
subalgebra in the full symmetry algebra of the absolute BRST cohomology of
the model.  After introducing $\CD$ by
$$\CD\equiv {}-\CG_{0,0},\eqn\newbrstcharge$$
and specific linear combinations $\CW_{m,n},\CW_a,\CW_b$ of $\CL^a_{m,n}$ and
$\CL^b_{m,n}$ by
$$\eqalign{\CW_{m,n}&\equiv n\CL^a_{m,n}-m\CL^b_{m,n},\cr
\CW_a\equiv&\,\CL^a_{0,0},\qquad \CW_a\equiv\CL^a_{0,0},\cr}
\eqn\winftycharges$$
we obtain the following commutation relations for the symmetry algebra of the
semirelative cohomology,
$$\eqalign{[\CW_{m,n},\CW_{p,q}]&=(mq-np)\CW_{m+p,n+q},\cr
[\CW_a,\CW_{m,n}]&=m\CW_{m,n},\cr [\CW_{m,n},\CQ_{p,q}]&=(mq-np)
\CQ_{m+p,n+q},\cr
[\CW_a,\CQ_{m,n}]&=m\CQ_{m,n},\cr\{\CD ,\CQ_{m,n}\}&=\CW_{m,n},\cr
[\CD ,\CW_a]&=0,\cr}
\qquad \eqalign{[\CW_a,\CW_b]&=0,\cr [\CW_b,\CW_{m,n}]&=n\CW_{m,n},\cr
\{ \CQ_{m,n},\CQ_{p,q}\}&=0,\cr [\CW_b,\CQ_{m,n}]&=n\CQ_{m,n},\cr
[\CD ,\CW_{m,n}]&=0,\cr [\CD ,\CW_b]&=0.\cr}
\eqn\winftytopnew$$
(In \winftytopnew , I have implicitly set $\CW_{0,0}\equiv\CQ_{0,0}
\equiv 0$.)

This algebra also has a nice spacetime interpretation.  The first two lines of
\winftytopnew\ form precisely $\winfty$, the algebra of all area-preserving
diffeomorphisms on the toroidal spacetime.  The preserved area is given by
$$\omega_{\rm area}=\frac{\der a}{a}\wedge\frac{\der b}{b}\equiv\der A
\wedge\der B,\eqn\areaform$$
and the explicit action of $w_\infty$ on the spacetime manifold $M$ is
realized by the following vector fields:
$$\eqalign{
\CW_{m,n}&=-n\, a^{m+1}b^n\,\frac{\p}{\p a}+m\, a^mb^{n+1}\,\frac{\p}{\p b},
\cr
\CW_a&= a\,\frac{\p}{\p a}, \quad\qquad\qquad \CW_b=b\,\frac{\p}{\p b}.\cr}
\eqn\winfaction$$
The $\CW_{m,n}$ vector fields of \winfaction\ correspond to the
area-preserving spacetime diffeomorphisms generated locally by Hamiltonians,
while the remaining two generators $\CW_a$ and $\CW_b$ correspond to those
diffeomorphisms that are not generated by Hamiltonians, and are in one-to-one
correspondence with the generators of the first cohomology group of the
spacetime.  The algebra of area-preserving diffeomorphisms of a torus has been
found in [\arnold] and studied in the context of membrane physics in
[\winftytorus].

This establishes the existence of $\winfty$ symmetries in the ghost-zero part
of the semirelative BRST cohomology as defined by $G_0^-$.  Now we will
analyze the fermionic extension of this $\winfty$ algebra by the generators of
non-zero ghost numbers.

%%%%%%%%%%%%%%%%%%%%%%%%%%%%%%%%%%%%%%%%%%%%%%%%%%%%%%%%%%%%%%%%%%%%%%%%%%%%%
\section{Topological $\winfty$ on the Worldsheet and in the Spacetime}

The whole symmetry algebra \winftytopnew\ represents a simple fermionic
extension of $w_\infty$, and the vector-field representation of the symmetry
charges acquires a non-trivial dependence on $\Theta_a,\,\Theta_b$.  Using the
definition of $\CW$'s, eqn.\ \winftycharges , and the vector representation of
the original symmetry charges, eqn.\ \vectrepresent , the vector field
representation \winfaction\ gets completed to
$$\eqalign{
\CW_{m,n}&=-n\, a^{m+1}b^n\,\frac{\p}{\p a}+m\, a^mb^{n+1}\,\frac{\p}{\p b}
+a^mb^n(n\Theta_a -m\Theta_b)\left( m\frac{\p}{\p\Theta_a}+n\frac{\p}{\p
\Theta_b}\right) ,\cr
\CQ_{m,n}&=a^mb^n\left( m\frac{\p}{\p\Theta_a}+n\frac{\p}{\p\Theta_b}\right) ,
\qquad\qquad\ \ \CW_a= a\frac{\p}{\p a},\cr
\CD&\equiv {}-\CG_{0,0}=b\Theta_a\frac{\p}{\p b}-a\Theta_b\frac{\p}{\p a},
\qquad\qquad\CW_b=b\frac{\p}{\p b}.\cr}\eqn\winftopnewaction$$
The fermionic generators $\CQ_{m,n}$ in \winftytopnew\ transform with
respect to the diffeomorphism part of the algebra as modes of a two-tensor,
while $\CD$ tranforms as the zero mode of a vector.  This is precisely the
situation encountered in topologically twisted \ntwo\ superalgebras, so we
have arrived at another central result of the paper:  The full spacetime
symmetry algebra \winftytopnew\ in the semirelative BRST cohomology of the two
dimensional topological string theory is a topologically twisted \ntwo\
$\winfty$ superalgebra (denoted by $\winftytop$ throughout the paper).
\foot{Analogous $\winftytop$ algebras, as well as their $W^{\rm top}_\infty$
conterparts, were first studied in a different context (and with the
underlying manifold being a two-sphere) in [\winftop].}
$\CD$ is the BRST-like charge of this spacetime topological symmetry algebra,
and carries ghost number one on the worldsheet.  No central extension of
\winftytopnew\ results from the computation of the commutation relations by
the OPEs of the corresponding conserved currents on the worldsheet
[\toptorus], and none is actually allowed on general algebraic grounds
(cf.\ [\winftyrevnew]).

It is interesting to note as an aside remark that the topological $W_\infty$
algebra of the sphere was first realized in [\winftop] as the algebra of {\it
worldsheet} symmetries in the theory described by our Lagrangian \ttlag ;
indeed, the topological Virasoro symmetry \wstopovirasoro\ actually extends to
the topological $W_\infty$ algebra.  In terms of the free fields that enter
the Lagrangian, the (left-moving) currents of the topological
$W^{\rm top}_\infty$ superalgebra of worldsheet symmetries are given by
[\winftop]
$$\eqalign{T^j(z)&\equiv\sum_m\frac{W^j_m}{z^{m+2}}\cr
&=(-1)^j\frac{2^j (j+1)!}{(2j+1)!!}\sum_{k=0}^j\binom jk \binom{j+1}k
(\p^{j-k+1}X\p^{k+1}\bar X+\p^{j-k+1}\chi\p^k\psi ),\cr
G^j(z)&\equiv\sum_m\frac{G^j_m}{z^{m+2}}=
(-1)^j\frac{2^j (j+1)!}{(2j+1)!!}\sum_{k=0}^j\binom jk \binom{j+1}k
\p^{j-k+1}X\p^k\chi ,\cr
Q(z)&\equiv \sum_m\frac{Q_m}{z^{m+1}}= \psi\p\bar X.\cr}
\eqn\wswswsws$$
For $j=0$ we recover the topological Virasoro superalgebra of \wstopovirasoro\
and \wsviracurrents .

The full $W^{\rm top}_\infty$ algebra of the worldsheet charges can be found
in [\winftop].  Here we just note that the $W_\infty^{\rm top}$ algebra can be
contracted to the corresponding $\winftytop$ algebra, in the classical limit
of the worldsheet CFT.  Commutation relations of this classical $\winftytop$
algebra are given by
$$\eqalign{[W^j_m,W^k_n]&=\{ (k+1)m-(j+1)n\}\,W^{j+k}_{m+n},\cr
[W^j_m,G^k_n]&=\{ (k+1)m-(j+1)n\}\,G^{j+k}_{m+n},\cr\{ G^j_m,Q_0\}&=W^j_m,\cr
}\qquad\eqalign{[W^j_m,Q_0]&=0,\cr\{ G^j_m,G^k_n\}&=0,\cr\{ Q_0,Q_0\}&=0,\cr}
\eqn\comroldold$$
and can be contrasted with the commutation relations of the $\winftytop$
algebra \winftytopnew\ that we have obtained in spacetime.

To conclude this aside remark on worldsheet symmetries of the model, we have
seen that, as a result of [\winftop], the worldsheet topological Virasoro
algebra of the topological torus is extended to the full quantum
$W^{\rm top}_\infty$ on the worldsheet, whereas as one of the cetral results
of the present paper, the same model also enjoys (in the semirelative BRST
cohomology) a topological $w_\infty$ symmetry in spacetime.  There are
obvious differences between the roles played by the worldsheet and spacetime
$\winfty$ symmetries in our model.  While the worldsheet $W^{\rm top}_\infty$
is a symmetry of the full Hilbert space of the worldsheet CFT, the spacetime
$\winftytop$ symmetry maps physical states to physical states.  The form of
the currents \wswswsws\ of the worldsheet $W^{\rm top}_\infty$ symmetry is
also very different from our expressions for the currents of the topological
$\winfty$ symmetry in spacetime.  On the other hand, in string theory we
generically expect that quantum corrections from higher genera deform
spacetime $\winfty$ symmetries to their $W_\infty$ counterparts; if this
happens with the spacetime $\winfty$ symmetry of the topological theory
studied here, it will strengthen an interesting similarity between its
worldsheet and spacetime symmetries.

%%%%%%%%%%%%%%%%%%%%%%%%%%%%%%%%%%%%%%%%%%%%%%%%%%%%%%%%%%%%%%%%%%%%%%%%%%%
\section{Semirigid $\winfty$ Geometry in Spacetime}

We have demonstrated that the spacetime symmetry algebra \winftytopnew\ of the
semirelative BRST cohomology, as generated by conserved worldsheet charges,
forms a $\winfty$ analog of the topologically twisted \ntwo\ Virasoro
algebra.  However, the usual topological Virasoro superalgebra in two
dimensions contains, besides the bosonic Virasoro generators and their BRST
fermionic superpartners, an infinite number of modes of the BRST current.  In
our case, we have obtained bosonic $\winfty$ generators $\CW_{m,n}$, their
superpartners $\CQ_{m,n}$, and the nilpotent fermionic charge $\CD$.  This
fermionic charge is the zero mode of the spacetime BRST-like current, and one
might wonder why we haven't also found the non-zero modes of the current.

The fact that just the zero mode of the spacetime BRST current emerges in the
topological symmetry algebra is probably not so surprising; rather it is
reminiscent of symmetries possessed by two dimensional topological gravity.
Indeed, precisely the same pattern, with just the zero mode of the BRST
current accompanying the bosonic symmetries and their BRST superpartners, has
emerged in semirigid geometry [\nelson], which is the suitable geometrical
framework for topological gravity.  Instead of leading to the full
topologically twisted \ntwo\ algebra, semirigid geometry ends up naturally
with the algebra consisting of the topological BRST charge, the Virasoro
algebra $L_m$, and the superpartners $G_m$ of the Virasoro generators $L_m$
under the BRST charge; the rest of the topologically twisted \ntwo\ Virasoro
superalgebra, in particular the higher modes of the BRST current, is broken.

In our case, the symmetry algebra of the topological torus induces naturally
what might be called a ``semirigid $\winfty$ structure'' on spacetime.
In view of the relationship between the absolute and semirelative BRST
cohomologies of the model, the semirigid $\winfty$ structure on $M$ can be
obtained by introducing the $\Delta$ operator and completing thus the
odd-symplectic structure on $\CM$ to the full Batalin-Vilkovisky geometry.
The $\Delta$ operator induces a volume element on $M$, and allows one
to identify canonically $TM$ with the cotangent bundle $T^*M$.  The geometry
of the $\winfty$ symmetries can then be formulated in terms of geometrical
structures on $T^*M$, so that we find a close contact with recent results in
$W$ geometry [\hull,\wsurp].  To see whether the relationship between the
odd-symplectic and $W$ geometries can be of some further significance, it
would be necessary to extend the results sketched here to surfaces of higher
genera, which is however beyond the scope of this paper.

%%%%%%%%%%%%%%%%%%%%%%%%%%%%%%%%%%%%%%%%%%%%%%%%%%%%%%%%%%%%%%%%%%%%%%%%%%%%
%
\chapter{Spontaneous Breakdown of Spacetime Diffeomorphisms}
\chaptermessage{Spontaneous Breakdown of Spacetime Diffeomorphisms}

The Lagrangian of the topological torus \ttlag\ has its own class of possible
deformations that preserve the topological BRST invariance:  any two-form that
enters the descent equations for the BRST charge can be formally used as an
additional term in the Lagrangian without spoiling the topological symmetry.
The deformed model then realizes only a part of the original huge spacetime
symmetry, and the rest of the symmetry algebra is spontaneously broken.

The basic strategy is as follows.
\foot{Here I follow closely the analogous discussion carried out for the
dynamical phase of two dimensional string theory in [\vermaster].}
Assume that we deform the Lagrangian by a two-form $\CO^{(2)}$, BRST invariant
up to exterior derivative of $\CO^{(1)}$,
$$I(\alpha )=I_0+\alpha\int_\Sigma O^{(2)}.\eqn\sampledef$$
Here $\alpha$ is a coupling constant.  The BRST charge of the original model
gets also deformed, to
$$Q(\alpha )=Q+\alpha\oint_C\CO^{(1)}.\eqn\samplebrst$$
Physical observables of the deformed theory belong to the BRST cohomology of
$Q(\alpha )$, so that they satisfy
$$Q|{\rm phys}\rangle=-\alpha\oint_C\CO^{(1)}\;|{\rm phys}\rangle .
\eqn\samplephys$$
In view of \samplephys , only those symmetry charges of the undeformed model
that commute with the charge $\oint\CO^{(1)}$ will survive, and the rest of
the original symmetry algebra may be spontaneously broken.  Of course, to
establish this result, one must show that no other symmetry charges are
generated that were not present in the the original symmetry algebra of the
undeformed model.

In this section we first study deformations of the theory within the absolute
BRST cohomology, and return to the semirelative BRST cohomology in \S{5.4}.
The results of the perturbative analysis of the symmetry brakdown will be
rather formal, as the method ignores the analogy of the disappearance and
re-appearance of some of the discrete modes as a result of the target
kinematics; the proper interpretation of the results would require a wider
framework which would allow us to consider all possible modes at the same
time, possibly similar to the framework proposed in the context of \ntwo\
superstrings in [\shapere].

%%%%%%%%%%%%%%%%%%%%%%%%%%%%%%%%%%%%%%%%%%%%%%%%%%%%%%%%%%%%%%%%%%%%%%%%%%%%%
\section{Topological Deformations and $\winfty$ Diffeomorphisms}

There is one natural candidate for the deforming two-form, carrying zero
winding numbers as well as zero ghost number.  This two-form is the top
element of the descent equation for $R^{(0)}_{0,0}\equiv \psi\bar\psi$.
Note that $R^{(0)}_{0,0}$ is a pure ``homology observable'' in the terminology
of [\toptorus], i.e.\ it does not contain contributions coming from the
non-zero fundamental group of the target, and would be present even if the
fundamental group were zero.  This class of observables in topological sigma
models is much better understood than generic observables which get
contributions from the non-zero fundamental group, so in this subsection
we can essentially borrow results from the theory of simply-connected
topological sigma models.

On-shell the two-form that will be used as a deformation of \ttlag\ is given
by
$$R^{(2)}_{0,0}(z,\bar z)\equiv \p X\wedge \bar\p \bar X .\eqn\defterm$$
After being integrated over a compact worldsheet $\Sigma$, $R^{(2)}_{0,0}$
becomes BRST invariant, and can deform the Lagrangian to
$$I(\alpha_{0,0})=I_0+\alpha_{0,0}\int_\Sigma R^{(2)}_{0,0}(z,\bar z).
\eqn\deflagtrin$$
The new term in the Lagrangian (unlike its analogs with non-zero winding
numbers) has a natural geometrical interpretation:  it essentially measures
the element of the second homology group spanned by the mappings of $\Sigma$
to the target.

In the deformed theory, the symmetry algebra is broken to the algebra that
respects the new term in the Lagrangian.  This condition can be usefully
formulated in terms of the symmetry algebra itself:  Via the topological
descent equations, $R^{(2)}_{0,0}$ defines a conserved charge
$$\CG_{0,0}\equiv \frac{1}{2\pi}\oint _C R^{(1)}_{0,0}(z,\bar z),
\eqn\defocharge$$
and non-zero $\alpha_{0,0}$ deforms the BRST charge $Q$ to
$$Q\rightarrow Q+\alpha_{0,0}\,\CG_{0,0}.\eqn\deformedbrst$$
The unbroken subalgebra of \communrelunbr\ consists of those charges that
commute with the new BRST charge of \deformedbrst , and hence with
$\CG_{0,0}$.

On the ground ring, the fermionic charge given by \defocharge\ acts as
$$\CG_{0,0}= a\Theta_b\,\frac{\p}{\p a}-b\Theta_a\,\frac{\p}{\p b}.
\eqn\defocharact$$
Among the $\CQ_{m,n}$'s, just $\CQ_{0,0}$ commutes formally with $\CG_{0,0}$,
but we already know from [\toptorus] that $\CQ_{0,0} \equiv 0$ identically.
In the rest of the fermionic sector, all $\CG_{m,n}$ survive, since they carry
the top ghost number and must anticommute with one another, in particular with
$\CG_{0,0}$.

In the bosonic sector, we have
$$[\CL^a_{m,n},\CG_{0,0}]=m\,\CG_{m,n},\qquad
[\CL^b_{m,n},\CG_{0,0}]=n\,\CG_{m,n},\eqn\bosecomzero$$
hence the subalgebra of bosonic charges commuting with $\CG_{0,0}$
consists of combinations
$$\CW_{m,n}\equiv n\CL^a_{m,n}-m\CL^b_{m,n}\eqn\winftycharg$$
and
$$\CW_a\equiv \CL^a_{0,0},\qquad \CW_b\equiv \CL^b_{0,0}.\eqn\transwinf$$
The algebra of all survivors is then
$$\eqalign{[\CW_{m,n},\CW_{p,q}]&=(mq-np)\CW_{m+p,n+q},\cr
[\CW_a,\CW_{m,n}]&=m\CW_{m,n},\cr [\CW_{m,n},\CG_{p,q}]&=(mq-np)\CG_{m+p,n+q},
\cr
[\CW_a,\CG_{m,n}]&=m\CG_{m,n},\cr}
\qquad \eqalign{[\CW_a,\CW_b]&=0,\cr [\CW_b,\CW_{m,n}]&=n\CW_{m,n},\cr
\{ \CG_{m,n},\CG_{p,q}\}&=0,\cr [\CW_b,\CG_{m,n}]&=n\CG_{m,n}.\cr}
\eqn\winftyalg$$

The first two lines of \winftyalg\ are again the $\winfty$ algebra of all
area-preserving diffeomorphisms of the toroidal spacetime, identified in the
previous section as the full algebra of ghost-number-zero symmetries in the
semirelative BRST cohomology.  Here we have just seen that the algebra of
spacetime diffeomorphisms, which is the algebra of ghost-number-zero
symmetries in the absolute BRST cohomology, can be broken spontaneously to a
$\winfty$ subalgebra.

Now we are going to analyze the impact of the spontaneous symmetry breakdown
on the full symmetry algebra of all ghost numbers.  The whole algebra
\winftyalg\ is a fermionic extension of $w_\infty$, and the symmetry charges
act on the spacetime supermanifold by the following vector fields,
$$\eqalign{
\CW_{m,n}&=-n\, a^{m+1}b^n\,\frac{\p}{\p a}+m\, a^mb^{n+1}\,\frac{\p}{\p b}+
a^mb^n(n\Theta_a -m\Theta_b)\left( m\frac{\p}{\p\Theta_a}+
n\frac{\p}{\p\Theta_b}\right) ,\cr
&\CW_a= a\frac{\p}{\p a}, \qquad\qquad\qquad \CW_b=b\frac{\p}{\p b},\cr
&\CG_{m,n}=a^{m+1}b^n\Theta_b\frac{\p}{\p a}-a^mb^{n+1}
\Theta_a\frac{\p}{\p b}+a^mb^n\Theta_a\Theta_b\left( m\frac{\p}{\p\Theta_a}+
n\frac{\p}{\p\Theta_b}\right) .\cr}\eqn\winftopaction$$
Their commutation relations \winftyalg\ are again those of a $\winftytop$
superalgebra.  The role of the fermionic superpartners of $\CW_{m,n}$ is now
played by charges $\CG_{m,n}$ of ghost number one, rather than by charges
$\CQ_{m,n}$ of ghost number minus one as in \S{4}.

%%%%%%%%%%%%%%%%%%%%%%%%%%%%%%%%%%%%%%%%%%%%%%%%%%%%%%%%%%%%%%%%%%%%%%%%%%%%%
\section{$\Delta$ as the BRST-like Charge in the Unbroken $\winftytop$ Algebra}

We have seen above that the symmetry algebra of unbroken charges follows the
general pattern of topological symmetry algebras, related to \ntwo\
superalgebras by twisting.  The only missing ingredient that would complete
\winftyalg\ to the topologically twisted \ntwo\ $\winfty$ superalgebra is
the scalar fermionic supercharge, whose anticommutation relations with the
fermionic generators create the bosonic ones.  In the standard interpretation
of topological superalgebras, this solitary fermionic charge plays the role of
the BRST charge.

To be more specific, we are looking for a fermionic charge $\BD$ with
commutation relations
$$\{ \BD ,\CG_{m,n}\} =\CW_{m,n},\qquad [\BD ,\CW_{m,n}]=0
\eqn\commsttop$$
(where I implicitly set $\CW_{0,0}\equiv 0$).  It is also natural to require
that $\BD$ commute with the two exceptional bosonic generators of the
symmetry algebra,
$$[\BD ,\CW_a]=0,\qquad [\BD ,\CW_b]=0.\eqn\excepdeecr$$
One can easily see that even the commutation relations \commsttop\ themselves
cannot be satisfied with $\BD$ a first order differential operator in
$a,\, b,\, \Theta_a,\,\Theta_b$.  Instead, they are satisfied by $\BD$
being a {\it second order} differential operator,
$$\BD =a\frac{\p^2}{\p a\,\p\Theta_a}+b\frac{\p^2}{\p b\,\p\Theta_b}.
\eqn\sttopsec$$
This is exactly the $\Delta$ operator \deltaop\ we encountered several times
in previous sections.  Consequently, the BRST-like charge $\BD$ of the
$\winftytop$ superalgebra \winftyalg\ can be identified with the BV $\Delta$
operator,
$$\BD\equiv\Delta ,\eqn\someidentity$$
which indicates an interesting relationship between the deformation of
the Lagrangian studied here, and the semirelative BRST cohomology studied in
\S{4}.

Note also that the existence of the BRST-like operator $\BD\equiv\Delta$
allows us to construct the symmetry algebra $\winftytop$ of the deformed model
by starting with the fermionic symmetries $\CG_{m,n}$ and generating the rest
of $w^{\rm top}_\infty$ by (anti)commutators with $\Delta$.  The only
symmetries that are not generated by this procedure are the non-exact
diffeomorphisms $\CW_a$ and $\CW_b$.

%%%%%%%%%%%%%%%%%%%%%%%%%%%%%%%%%%%%%%%%%%%%%%%%%%%%%%%%%%%%%%%%%%%%%%%%%%%%%
\section{Higher Deformations with Zero Ghost Number}

Instead of the translation-invariant two-form $R^{(0)}_{0,0}$, we can
use the whole hierarchy of two-forms coming from the descent equations, to
deform the original Lagrangian \ttlag\ and obtain a model with a spontaneous
breakdown of the original spacetime diffeomorphism symmetry.  For any
$N\in {\bf Z}$ we can study
$$I(\alpha_{N,N})=I_0+\alpha_{N,N}\int_\Sigma R^{(2)}_{N,N}(z,\bar z),
\eqn\deflagnonc$$
where $R^{(2)}_{N,N}$ is given by
$$R^{(2)}_{N,N}=(\p X -\ii k_{N,N}\psi\chi )\wedge (\bar\p\bar X+\ii
\bar k_{N,N}\bar\psi\bar\chi )\e{\ii k_{N,N}\bar X(z)-\ii\bar k_{N,N}
X(\bar z)},\eqn\highdeftwo$$
with
$$k_{m,n}=R(m+n\tau_0). \eqn\momenta$$
Note that $I(\alpha_{N,N})$ is complex; if one wants a real deformation of
the Lagrangian, one can deform $I_0$ by real linear combinations of the
deforming two-forms.

The residual bosonic symmetry generators of the deformed model are given by
$$\eqalign{W^{N}_{m,n}&\equiv (n-N)\CL^a_{m,n}-(m-N)\CL^b_{m,n},\cr
W^{N}_a&\equiv \CL^a_{N,N}, \qquad\qquad W^{N}_b\equiv \CL^b_{N,N},\cr}
\eqn\survnonc$$
and form the following symmetry algebra:
$$\eqalign{[W^{N}_{m,n},W^{N}_{p,q}]&=C^N_{mn,pq} W^{N}_{m+p,n+q},\cr
[W^{N}_a,W^{N}_{m,n}]&=(N-m)\, W^{N}_{m+N,n+N},
\cr
[W^{N}_b,W^{N}_{m,n}]&=(N-n)\, W^{N}_{m+N,n+N},\cr}\eqn\symalgnonc$$
with the structure constants $C^N_{mn,pq}$ given by
$$C^N_{mn,pq}\equiv(m-N)(q-N)-(n-N)(p-N).\eqn\strucconst$$
This algebra preserves a higher two-form on the torus,
$$\omega^{N}_{\rm area}=\frac{\der a}{a^{N+1}}\wedge\frac{\der b}{b^{N+1}}.
\eqn\sympformnonc$$

Among the fermionic generators, $\CG_{m,n}$ will again survive for all
$m,n$, and extend the bosonic symmetry algebra \symalgnonc\ to a topological
$w_\infty$.  The relevant commutation relations are
$$\eqalign{[W^{N}_{m,n},\CG_{p,q}]&=C^N_{mn,pq}\CG_{m+p,n+q},\cr
[W^{N}_a,\CG_{m,n}]&=(N-m)\,\CG_{m+N,n+N},
\cr
[W^{N}_b,\CG_{m,n}]&=(N-n)\,\CG_{m+N,n+N},\cr}\eqn\symalgnoncfer$$
with $C^N_{mn,pq}$ given again by \strucconst .  Note that the translational
invariance in spacetime is broken for non-zero $N$, since the generators of
spacetime translations $\CL^a_{0,0}$ and $\CL^b_{0,0}$ are no longer elements
of the unbroken symmetry algebra.

%%%%%%%%%%%%%%%%%%%%%%%%%%%%%%%%%%%%%%%%%%%%%%%%%%%%%%%%%%%%%%%%%%%%%%%%%%%%%
\section{Symmetry Breakdown in the Semirelative BRST Cohomology}

The BRST invariant two-form \defterm , used in \S{5.1} as a deformation of the
basic Lagrangian \ttlag , belongs to the semirelative BRST cohomology defined
by the BV $\Delta$ operator.  In view of this, we can study the deformation
by $R^{(2)}_{0,0}$ in the model with physical states defined by the
semirelative cohomology condition, \brstequiv .  The natural question is, how
is the $\winftytop$ superalgebra of spacetime symmetries broken when the model
is deformed by $R^{(2)}_{0,0}$.

To answer this question, note first that the conserved charge $\CG_{0,0}$
related to $R^{(2)}_{0,0}$ by the BRST descent equation, is nothing but the
spacetime BRST-like charge $\CD$ of the semirelative $\winftytop$
superalgebra.  In this sense, the model is deformed by its own BRST-like
spacetime symmetry charge!  The surviving symmetries are those that commute
with this BRST-like charge, leading to a symmetry algebra that contains just
the bosonic generators $\CW_{m,n}$ and the BRST-like charge $\CD\equiv
-\CG_{0,0}$ itself; the resulting commutation relations are
$$\eqalign{[\CW_{m,n},\CW_{p,q}]&=(mq-np)\CW_{m+p,n+q},\qquad [\CW_a,\CW_b]=0,
\cr [\CW_a,\CW_{m,n}]&=m\CW_{m,n},\qquad\qquad\qquad\quad [\CW_b,\CW_{m,n}]=
n\CW_{m,n},\cr[\CW_{m,n},\CG_{0,0}]&=0,\qquad [\CW_a,\CG_{0,0}]=0,\qquad
[\CW_b,\CG_{0,0}]=0.\cr}\eqn\intersectw$$
This is the unbroken symmetry algebra in the semirelative BRST cohomology.

To conclude this section, we have seen that it is possible to deform the model
of \S{4} by the spacetime BRST-like charge $\CD$ of the $\winftytop$ spacetime
superalgebra.  The only fermionic charge that survives in the semirelative
BRST cohomology is decoupled from the rest of the spacetime symmetry algebra,
which thus becomes the usual, bosonic $\winfty$.

%%%%%%%%%%%%%%%%%%%%%%%%%%%%%%%%%%%%%%%%%%%%%%%%%%%%%%%%%%%%%%%%%%%%%%%%%%%%%
%
\chapter{Concluding Remarks}
\chaptermessage{Concluding Remarks}

In this paper I have presented a model whose spacetime symmetries are
closely related to all spacetime diffeomorphisms, and have found a mechanism
which reduces the diffeomorphism symmetry to a $\winfty$ algebra.  This model
is a two dimensional topological string theory, and the residual unbroken
symmetry is quite remimiscent of the manifest $\winfty$ symmetry of the
dynamical phase of string theory in two dimensions.  We have also seen that
physical observables of higher ghost numbers extend the $\winfty$ symmetry
algebra to a spacetime topological $\winfty$ superalgebra, with its own
BRST-like nilpotent fermionic charge.  This suggests that the theory may be
cohomological not only on the worldsheet but also in spacetime, a conjecture
reminiscent of some other recent observations in topological string theory
[\elitzur,\wcswstrings].

The spacetime manifold, as defined by the generators of the ground ring, is
dual to the target manifold, as defined by the field content of the
sigma-model Lagrangian.  This duality between the spacetime and the target
suggests a mechanism capable of relating the topological phase of a physical
theory to its phase with local dynamics.  Indeed, the set of all point-like
physical states of the topological sigma model consists entirely of ground
states in each winding sector on the target.  Since as we have seen the target
is dual to the spacetime, these target winding modes can be (roughly)
ininterpreted as momentum modes in spacetime, and we find local degrees of
freedom that arise as topological physical states in a cohomological string
theory.  I believe that the lesson learned from this example might be of some
wider validity in topological field theory.

The model I have studied in this paper can be extended straightforwardly to
topological string theory on higher-dimensional tori of real dimension 2d.
Most of the results of this paper find their natural counterparts in higher
dimensions.  For example, the symmetries in the absolute BRST cohomology
generate all spacetime diffeomorphisms at ghost number zero, extended at
non-zero ghost numbers to the algebra of all odd-symplectic diffeomorphisms on
a spacetime supermanifold of dimension (2d$|$2d); in the semirelative BRST
cohomology, the ghost-number-zero symmetry algebra preserves a volume element
on spacetime.  There is, however, one important property that does not extend
to higher dimensions, and makes the two-dimensional case unique: The symmetry
algebra of the semirelative BRST cohomology cannot be interpreted as a
topologically twisted \ntwo\ superalgebra unless the spacetime dimension is
two.

Another possible generalization is represented by topological string theory on
higher genus Riemann surfaces, which is interesting from several points of
view.  First, if one expects summation over all spacetime topologies to enter
the full string theory, one would face in this particular case the problem of
formulating the topological sigma model on surfaces with higher genera.
The topological sigma model with such targets can indeed be formulated, and is
clearly not conformally invariant.  This elementary fact prevents us from
applying most of the methods we have used for the target torus, and would
probably require much better understanding of general topological sigma models
with multiply connected targets.  A better understanding of this theory would
be of some wider importance in various areas of physics.

Two dimensional string theory from the topological point of view has recently
been studied by Mukhi and Vafa [\mukhivafa].  Their results are extremely
interesting, and complementary to ours; using the insight of [\newtopo], the
authors of [\mukhivafa] present a strong evidence showing that there is a
hidden topological symmetry on the worldsheet of the dynamical phase of
two dimensional string theory, in particular for the black hole solution.  In
one case of their interest, the authors of [\mukhivafa] have recovered the
Lagrangian \ttlag\ that serves as the starting point of the present paper.
It would be very interesting to combine their worldsheet results with the
spacetime results obtained in this paper.
\bigskip
\centerline{\fourteenpoint Note added}
\smallskip

Quite recently, the importance of odd-symplectic geometry and the BV formalism
in string theory and 2-D topological field theory has also been stressed
by Lian and Zuckermann; Penkava and Schwarz; and Getzler [\newbvpapers].
\bigskip
\centerline{\fourteenpoint Acknowledgements}
\smallskip

It is a pleasure to thank A. Ashtekar, J. Avan, R. Dijkgraaf, J. Distler,
P. Nelson, A. Shapere, J. Stasheff, E. Verlinde, E. Witten and B. Zwiebach for
stimulating discussions at various stages of the work.  The author is grateful
to the organizers of the \'Ecole d'\'Et\'e de Physique Th\'eorique on
``Gravitation and Quantizations'' (July 1992) for their hospitality in Les
Houches, where part of this work was done.  The author is also indebted to the
organizers of the JAMI conference on ``Geometry and Quantum Field Theory,''
Baltimore (March 1992); Spring School on String Theory and Quantum Gravity,
Trieste (April 1992); and the \'Ecole d'\'Et\'e in Les Houches (July 1992),
for the opportunity to present various aspects of these results.

\message{* References}
\refout
\end
\bye